\documentclass{article}
\usepackage{graphicx}
\usepackage{latexsym}
\newcommand{\be}{\begin{equation}}
\newcommand{\ee}{\end{equation}}
\newcommand{\bea}{\begin{eqnarray}}
\newcommand{\eea}{\end{eqnarray}}

\newcommand{\ovl}{\overline}

\begin{document}

\title{\bf Kerr spacetime in Lema\^itre coordinates}

\author{
Francesco Sorge\\ 
{\it I.N.F.N. - Compl. Univ. di Monte S. Angelo,}\\
{\it via Cintia, Edificio 6,  I-80126 Napoli, Italy}\\
{\rm sorge@na.infn.it}}

\maketitle

\begin{abstract}
We present a novel formulation of the Kerr spacetime solution, based on the Lema\^itre coordinates. Such an approach allows one to avoid the coordinate singularities of the Boyer-Lindquist metric, thus offering the possibility to explore in some detail the Kerr physics near and beyond the event horizon, adopting the point of view of a freely falling observer, whose adapted tetrad is moreover proven to be Fermi-Walker transported.  We use the Lema\^itre form of the Kerr metric discussing the motion of radial photons close to the Cauchy horizon, also showing that their behaviour fairly agrees with other similar results presented in the literature.  Being synchronous, the Lema\^itre form of the Kerr metric allows a Hamiltonian formulation of a quantum field theory nearby and beyond the Kerr event horizon. This could provide a further tool in the investigation  of the intriguing issue concerning the Cauchy horizon (in)stability.
\end{abstract}

pacs: 04.20.-q,  04.20.Jb

%%%%%%%%%%%%%%%%%%%%%%%%%%    section 1     %%%%%%%%%%%%%%%%%%%%%%%%%%%%%%%%%%
\section{Introduction}

Since its discovery in 1963, the Kerr solution \cite{kerr} of the vacuum Einstein field equations has been deeply investigated as one of the most promising and useful solutions in the study of stellar collapse, as well as of black hole formation and evolution. 
Kerr spacetime admits two horizons: an outer, {\em event} horizon (EH), and an inner horizon, the {\em Cauchy} horizon (CH). The internal physical {\em ring} singularity is hidden behind the event horizon, hence saving us from direct comparison with experimental data, meanwhile we are lacking a satisfactory theory of gravity at the Planck scale. However, well before we have to tackle the hidden singularity, we need to face an open - and more severe - issue, concerning the stability of the Cauchy horizon against perturbations, both at the classical and quantum level. Such an  issue could lead to the breakdown of General Relativity itself in some peculiar circumstances, since Cauchy horizons are a feature of several spacetime solutions of the vacuum Einstein field equations.

For example, in the case of the Reissner-Nordstr\"om-de Sitter spacetime, it is widely believed that any perturbation should blow up badly enough when approaching the Cauchy horizon \cite{chandra2,toporenski}, thus rendering the latter unstable at least in the {\em quantum} limit \cite{hollands}. On the other hand, there is some evidence that {\em classical} scalar perturbations remain finite, causing no relevant pathology at the Cauchy horizon and eventually violating the so-called {\em strong cosmic censorship conjecture} \cite{cardoso1,cardoso2}.

Given the above controversy, further investigation of the behaviour of both classical and quantum fields nearby the Cauchy horizon of a Kerr black hole seems mandatory. Various coordinate sets can be employed in order to highlight the very many properties of the Kerr spacetime \cite{chandra,visser}. In carrying on such analysis a coordinate set {\em regular} at the horizon is required.

Historically, the first coordinate set regular at $r =2M\equiv r_g$  was obtained, independently, by P. Painlev\'e \cite{painleve} and A. Gullstrand \cite{gullstrand}, considering the Scwharzschild spacetime. Other coordinate sets have been found afterwards. We recall the Eddington-Finkelstein coordinates \cite{eddington,finkelstein}, adapted to radial null geodesics (e.g., photons) and the Kruskal-Szekeres coordinates \cite{kruskal}, which allow to explore the maximal analytic extension of the Schwarzschild solution. 

Another interesting coordinate set was proposed by G. Lema\^itre, who first recognized in 1933 \cite{lemaitre} that the Schwarzschild singularity at the horizon $r = r_g$ is a mere {\it coordinate} singularity. Beyond the horizon, standard Schwarzschild coordinates cannot be described by means of static material bodies. Actually, when $r < r_g$ the radial coordinate behaves like a {\em time} coordinate, so that particles cannot be at rest at a constant radius $r$. 
On the other hand, Lema\^itre coordinates, obtained via a suitable transformation of the standard Schwarzschild coordinates \cite{kramer}, are adapted to radial {\em time-like} geodesics. In other words, such coordinates are comoving with freely falling massive test particles (or observers), so that the latter have a {\em constant} value of the radial coordinate during the fall. As a consequence, Lema\^itre coordinates are well-behaved at the horizon $r=r_g$. 

Turning back to the more involved geometry of the Kerr spacetime, a new coordinate set, free from singularities at the horizons and closely related to the Painlev\'e-Gullstrand coordinates, was proposed by  Doran \cite{doran} about twenty years ago.
In his approach Doran obtained a new form for the Kerr metric by means of a suitable transformation of the advanced Eddington-Finkelstein coordinates. The resulting {\em non-diagonal} metric is well-behaved at the horizons, hence allowing to analyze the physics of Kerr black hole near the event horizon and beyond it.  Also, the time coordinate $t$ coincides with the proper time of an observer in radial free fall.

In the present paper we will propose a novel formulation of the Kerr metric, based on a modified form of the Lema\^itre transformation applied to the spheroidal oblate Boyer-Lindquist coordinates.  
Due to the analogy with the original Lema\^itre coordinate set (first employed in the Schwarzschild spacetime), throughout the paper we will call for short {\em Lema\^itre coordinates}  the new coordinates in which we will recast the Kerr metric.

Let us summarize a few interesting properties of the new coordinates:
\begin{itemize}
\item Lema\^itre coordinates are {\em horizon penetrating}, thus offering the possibility to explore the black hole physics beyond the event horizon, also by means of numerical techniques;
\item being adapted to time-like geodesics, such coordinates can be employed to describe the {\em physical point of view} of an infalling observer (or a massive test particle);
\item all the (freely falling) {\em Lema\^itre observers} experience the {\em same} proper time $\tau$, coincident with the {\em global} coordinate time.
\end{itemize}

We will also show that the resulting {\em Lema\^itre} form of the Kerr metric has several interesting features. In particular:
\begin{itemize}
\item it is synchronous (just as in the case of the Doran coordinates) and diagonal;
\item it is well-behaved at the horizons;
\item it reduces to the well-known Lema\^itre form of the Schwarzschild solution in the limit $a\rightarrow 0$;
\item thanks to its synchronous form, the obtained metric admits a Hamiltonian formulation of a quantum field theory \cite{fulling}. Hence  Lema\^itre  coordinates might prove useful to investigate the behaviour of {\em quantum fields} in the Kerr spacetime,  with particular care to the open issue concerning the Cauchy horizon (in)stability.
\end{itemize}

The paper is organized as follows. In section 2 we briefly recall the transformation between the standard Schwarzschild coordinates and the Lema\^itre coordinates, also writing the form of the Schwarzschild metric in the new coordinates. In section 3 we give a concise review of the Kerr metric, written in the popular Boyer-Lindquist coordinates.
In section 4 we obtain the transformation between the Boyer-Lindquist coordinates and a set of coordinates adapted to a freely-falling, zero-angular-momentum (ZAMO) observer \cite{bini}, representing the generalization of the Lema\^itre coordinates to the Kerr geometry. In section 5 we discuss the Kerr horizons in the new coordinate set, and focus on the family of Lema\^itre (i.e., freely falling) observers. In section 6 we analyze an imaginary journey towards the Cauchy horizon, considering in some detail the behaviour of radial photons excanged by two Lema\^itre observers. In section 7 we show that photon detection performed by Lema\^itre observers near the Cauchy horizon results in an infinite blue-shift. This is suggestive of Cauchy horizon instability \cite{chandra2,toporenski}, at least at the classical level. Finally, section 8 is devoted to some concluding remarks.

Throughout the paper, unless otherwise specified, use has been made of natural units:  $c=1$, $\hbar=1$, $G=1$. Greek indices take values from 0 to 3; latin ones take values from 1 to 3. The metric signature is $-2$, with determinant $g$.

%%%%%%%%%%%%%%%%%%%%%%%%%%    section 2     %%%%%%%%%%%%%%%%%%%%%%%%%%%%%%%%%%
\section{Lema\^itre coordinates: an overview}

In this section we briefly recall the Lema\^itre form for the Schwarzschild solution. A detailed derivation of the results below can be found, e.g.,  in \cite{frolov,sorge1}. 

The Schwarzschild metric for a black hole of mass $M$ in the standard Schwarzschild coordinates $\{t,r,\theta,\phi\}$ reads
\be\label{schw}
ds^2= \bigg(1-\frac{r_g}{r}\bigg)dt^2-\bigg(1-\frac{r_g}{r}\bigg)^{-1}dr^2-r^2 d\Omega^2,
\ee
where $r_g=2M$ is the gravitational (Schwarzschild) radius of the black hole and $d\Omega^2=d\theta^2+\sin^2\theta d\phi^2$. In such coordinates, we meet a {\em coordinate} singularity at the horizon. To analyze in detail the the behaviour of a test body falling into a black hole, we need a chart which is regular at the horizon, so the form (\ref{schw}) of the metric is not suitable. 

Consider then a massive test particle, radially falling with 4-velocity $\bf u$ in the gravitational field of the black hole. Since  (\ref{schw}) admits a time-like Killing vector, $\vec X=\partial_t$, we have a conserved quantity along a time-like geodesic, namely $\vec X\cdot \bf u=\gamma=$const, with $\gamma={\cal E}/m$ being the total specific energy of the test body (if the test particle starts its fall from rest at the spatial infinity, then $\gamma=1$).
Since for a radial infall motion $d\theta =d\phi=0$, we get from the constraint $g_{\mu\nu}u^\mu u^\nu=1$
\be\label{ur}
\frac{dr}{d\tau}=-\sqrt{\gamma^2-1+\frac{r_g}{r}},
\ee
where the sign refers to the radial {\em infall} and $\tau$ is the particle proper time. Notice that (\ref{ur}) implies
\be\label{vincle}
\gamma^2-1+\frac{r_g}{r}>0,
\ee
otherwise we have no radial motion. Such a constraint defines the allowed radial region as a function of the gravitational radius as well as the total specific energy of the falling particle.  Let us now introduce the following coordinate transformation (see \cite{kramer,sorge1})
\be
\left\{
	\begin{array}{ll}
dt= \frac{\gamma r}{r-r_g}d\tau+\bigg(\frac{1}{\gamma}-\frac{\gamma r}{r-r_g}\bigg)d\rho\nonumber\\
dr= \sqrt{\gamma^2-1+\frac{r_g}{r}}(-d\tau+d\rho).
\end{array}
\right.\label{transf}
\ee
Using (\ref{transf}) in (\ref{schw}) yields the Schwarzschild metric in the so-called {\em generalized} Lema\^itre coordinates $\{\tau,\rho,\theta,\phi\}$.
\be\label{glm}
ds^2=d\tau^2-\frac{1}{\gamma^2}\bigg(\gamma^2-1+\frac{r_g}{r(\tau,\rho)}\bigg)d\rho^2-r^2(\tau,\rho)d\Omega^2,
\ee
where the function $r(\tau,\rho)$ satisfies (\ref{ur}).

Suppose now a freely falling test particle with total specific energy $\gamma=1$. In this case, the Schwarzschild metric in the Lema\^itre coordinates reduces to
\be\label{lm}
ds^2=d\tau^2-\frac{r_g}{r(\tau,\rho)}d\rho^2-r^2(\tau,\rho) d\Omega^2.
\ee
Integrating (\ref{transf}) (along with $\gamma=1$), we obtain  the explicit relations between the Schwarzschild and the Lema\^itre coordinates \cite{frolov} 
\be\label{r}
r(\tau,\rho)=r_g^{1/3}\bigg[\frac{3}{2}(\rho-\tau)\bigg]^{2/3},
\ee
\be
t=r_g\Bigg[-\frac{2}{3}\left(\frac{r}{r_g}\right)^{3/2}-2\left(\frac{r}{r_g}\right)^{1/2}+\ln\bigg|\frac{(r/r_g)^{1/2}+1}{(r/r_g)^{1/2}-1}\bigg|+\frac{\rho}{r_g}\Bigg].
\ee

In so doing, we have introduced a {\em comoving coordinate} $\rho$, adapted to time-like geodesics: a test particle, moving along its geodesic has (at any proper time $\tau$) a {\em constant} value of the coordinate $\rho$. 
Thus, in the Lema\^itre coordinates, each freely falling observer is characterized by a {\em fixed} value $\rho$ of the radal coordinate.

Notice that the Lema\^itre coordinates define a {\em synchronous} frame, since the time coordinate $\tau$ is the proper time of all the comoving observers $\rho =\rm{const}$.

%%%%%%%%%%%%%%%%%%%%%%%%%%%%%%%%%%    section 3     %%%%%%%%%%%%%%%%%%%%%%%%%%%%%%%%%%
\section{Kerr metric in Boyer-Lindquist coordinates}
The Kerr solution is the only known family of exact, stationary, axisymmetric (with further reflection symmetry with respect to the equatorial plane), asymptotically flat solution, which is believed to represent the exterior space-time metric outside a rotating massive spherical object. The metric can be given in the Boyer-Lindquist coordinates ($t,r,\phi,\theta$), namely \cite{fdf}
\be\label{kerrbl}
 ds^2 =\bigg(1-\frac{2Mr}{\Sigma}\bigg)dt^2+\frac{4Mar}{\Sigma}\sin^2\theta dt d\phi
-\frac{\Sigma}{\Delta}dr^2-\Sigma d\theta^2-\frac{A}{\Sigma}\sin^2\theta d\phi^2,\label{bl}
\ee
where
\bea
\Sigma&=&r^2+a^2\cos^2\theta,\\
\Delta&=&r^2+a^2-2Mr,\\
A&=&(r^2+a^2)\Sigma+2Mra^2\sin^2\theta,
\eea
$M$ and $a=J/M$ are the mass and the specific angular momentum of the rotating body. The roots $r_\pm$ of $\Delta=0$ give the Kerr horizons, while those of $g_{tt}=0$ define the infinite redshift surfaces $r_{0\pm}$
\bea\label{horizon}
r_\pm&=&M\pm\sqrt{M^2-a^2},\\
r_{0\pm}&=&M\pm\sqrt{M^2-a^2\cos^2\theta}.
\eea
To avoid naked singularities we will assume $0<a<M$. In such a case, $r>r_+$ defines the physically accessible region, outside the event horizon $r_+$. The physical {\em ring} singularity ($r=0, \,\theta=\pi/2$) lies behind the Cauchy horizon, $r_-$.

%%%%%%%%%%%%%%%%%%%%%%%%%%%%%%%%    subsection 3.1     %%%%%%%%%%%%%%%%%%%%%%%%%%%%%%%%%%
\subsection{Introducing a rotating frame}
The metric (\ref{bl}) has off-diagonal terms, $g_{t\phi}\neq 0$. One can remove such terms introducing a suitable set of rotating coordinates
\be
\left\{
	\begin{array}{ll}
dt'= dt \nonumber\\
d\phi '= d\phi -\omega dt.
\end{array}
\right.\label{rotframe}
\ee
Using (\ref{rotframe}) we can diagonalize the metric (\ref{bl}), provided we choose
\be\label{omegadrag}
\omega = \frac{2Mar}{A}.
\ee
$\omega$ is the dragging angular velocity, namely the angular velocity at which the spacetime is dragged around the Kerr black hole.
The diagonalized form of the metric $g'_{\mu\nu}$ is
\be\label{bldiag}
ds^2=\frac{\Sigma \Delta}{A}dt'^2-\frac{\Sigma}{\Delta}dr^2 -\Sigma d\theta^2 -\frac{A}{\Sigma}\sin^2\theta d\phi'^2.
\ee

%%%%%%%%%%%%%%%%%%%%%%%%%%%%%%%%%%    subsection 3.2     %%%%%%%%%%%%%%%%%%%%%%%%%%%
\subsection{Symmetries and Killing vectors}
The independence of the Kerr metric of the coordinates $t$ and $\phi$ is a manifestation of the presence of two Killing vectors $\partial_t=\delta^\mu_t\partial_\mu$ and $\partial_\phi=\delta^\mu_\phi\partial_\mu$, satisfying the Killing equations
\be\label{killingeqs}
\nabla_\mu\xi_\nu+\nabla_\nu\xi_\mu=0.
\ee
The contraction of a Killing vector with the tangent vector $\dot\gamma^\mu$ of an affinely parametrized geodesic gives a conserved quantity, i.e., a constant of motion: $\xi^\mu\dot\gamma_\mu={\rm const}$.
In the case of a time-like geodesic (a massive test particle), we identify the tangent vector with the particle 4-velocity, $u^\mu\equiv\dot\gamma^\mu$.

Introducing the particle momentum $p^\mu=m u^\mu$ we thus find that the Kerr spacetime allows for four conserved quantities, in terms of which the geodesic motion of a test particle can be described. The first three are the particle mass $m$, the azimuthal angular momentum $L_z=-\delta^\mu_\phi=-p_\phi$ (i.e., the angular momentum with respect to the Kerr rotation axis, $\hat z$) and the energy $E=\delta^\mu_t p_\mu=p_t$. The fourth constant is the celebrated Carter's constant $\cal{K}$. One can recast these constants of motion introducing three specific conserved quantities, i.e., the specific angular momentum $\lambda=L_z/m$, the specific energy $\gamma=E/m$ and the specific Carter's  constant ${\cal{Q}}={\cal{K}}/m$
\bea\label{conserved}
\gamma=u_t=g_{t\mu}u^\mu=g_{t\mu}\frac{dx^\mu}{d\tau},\\
\lambda=-u_\phi=-g_{\phi\mu}u^\mu=-g_{\phi\mu}\frac{dx^\mu}{d\tau},
\eea
\be\label{carter}
{\cal{Q}}=\frac{\cal{K}}{m}=\frac{p_\theta^2}{m^2}+\bigg[a^2(1-\gamma^2)+\frac{\lambda^2}{\sin^2\theta}\bigg]\cos^2\theta,
\ee
where $\tau$ is the particle proper time along the geodesic.

%%%%%%%%%%%%%%%%%%%%%%%%%%%%%%%    subsection 3.3     %%%%%%%%%%%%%%%%%%%%%%%%%%%%%%%%%%
\subsection{Geodesic radial infall}

Consider a freely moving test particle, having mass $m$. In the Boyer-Lindquist coordinates (\ref{bl}) the geodesic motion is described by \cite{misner}
\bea\label{geodesic}
\quad\frac{dt}{d\tau}&=&\frac{A}{\Delta\Sigma}(\gamma-\lambda\omega),\\
\quad\frac{d\phi}{d\tau}&=&\frac{A}{\Delta\Sigma}\bigg[\bigg(\frac{\Sigma-2Mr}{A}\bigg)\frac{\lambda}{\sin^2\theta}+\omega\gamma\bigg],\\
\,\bigg(\frac{d\theta}{d\tau}\bigg)^2&=&\frac{1}{\Sigma^{2}}\bigg[{\cal{Q}}-\lambda^2\cot^2\theta+(\gamma^2-1)a^2\cos^2\theta\bigg],\\
\,\bigg(\frac{dr}{d\tau}\bigg)^2&=&\frac{1}{\Sigma^{2}}\bigg[\left(\gamma(r^2+a^2)-a\lambda\right)^2-\Delta\left(r^2+(\lambda-a\gamma)^2+{\cal{Q}}\right)\bigg],
\eea
If the particle is freely falling from the spatial infinity with zero initial velocity and zero {\em total} angular momentum, then $\gamma =1$, $\lambda =0$, and $p_\theta=0$. Hence ${\cal{Q}}=0$. So we have
\bea\label{geod}
\quad\frac{dt}{d\tau}&=&\frac{A}{\Delta\Sigma},\\
\quad\frac{d\phi}{d\tau}&=&\frac{2Mar}{\Delta\Sigma}=\frac{A}{\Delta\Sigma}\omega,\\
\quad\frac{d\theta}{d\tau}&=&0,\\
\,\bigg(\frac{dr}{d\tau}\bigg)^2&=&\frac{2Mr(\Delta+2Mr)}{\Sigma^2}=\frac{A-\Delta\Sigma}{\Sigma^2}.
\eea
The 4-velocity $\bf u$ of the infalling particle is then
\be\label{u}
{\bf u}=\frac{A}{\Delta\Sigma}\partial_t -\frac{\sqrt{A-\Delta\Sigma}}{\Sigma}\partial_r+\frac{A\omega}{\Delta\Sigma}\partial_\phi.
\ee
Hence the 4-velocity $\bf u'$ in the rotating frame $\{t',r,\theta,\phi'\}$, namely   
\be
u'^\mu = \frac{\partial x'^\mu}{\partial x^\alpha}u^\alpha,
\ee
reads
\be\label{uprimed}
{\bf u'}=\frac{A}{\Delta\Sigma}\partial_{t'} -\frac{\sqrt{A-\Delta\Sigma}}{\Sigma}\partial_r.
\ee

%%%%%%%%%%%%%%%%%%%%%%%%%%%%%%%%    section 4     %%%%%%%%%%%%%%%%%%%%%%%%%%%%%%%%%%
\section{Introducing the Lema\^itre coordinates}
We now move to the Lema\^itre coordinates. Consider the radial component of ${\bf u'}$ in the primed (rotating) frame $\{t',r,\theta,\phi'\}$ (hereafter, we will drop the primes for sake of clarity)
\be\label{rdot}
\frac{dr}{d\tau}=-\frac{\sqrt{A-\Delta\Sigma}}{\Sigma}.
\ee
Integrating we have
\be\label{rdotint}
-\int_c^\tau d\tau'=\int_{r_0}^r dr'\sqrt{\frac{\Sigma^2}{A-\Sigma\Delta}}=\frac{1}{\sqrt{2M}}\int_{r_0}^r dr'\frac{r'^2+a^2\cos^2\theta}{\sqrt{r'(r'^2+a^2)}},
\ee
where $c$ is a suitable integration constant.
Define the adimensional coordinate $z=r/a$ ($a\neq 0$). Then
\be\label{rdotint2}
-\int_c^\tau d\tau'=\frac{a^{3/2}}{\sqrt{2M}}\int_{z_0}^r dz'\frac{z'^2+\cos^2\theta}{\sqrt{z'(z'^2+1)}}.
\ee
The integral in the RHS can be solved in terms of elliptic functions of first kind, $F(z\,|m)$, namely
\be\label{tauint1}
-\int_c^\tau d\tau'=\frac{a^{3/2}}{\sqrt{2M}}\big[f(z,\theta)-f(z_0,\theta)\big],
\ee
where
\be\label{fzeta}
 f(z,\theta)=\frac{2}{3}\bigg[\sqrt{z(z^2+1)}+(-1)^{1/4}(3\cos^2\theta-1)F\left(i\,{\rm arc}\sinh \left(\frac{(-1)^{1/4}}{\sqrt{z}}\right)\bigg |-1\right)\bigg].
\ee
The function $f(z,\theta)$ is singular at $z=0$. However, it has a finite limit
\be\label{flimit}
 \lim_{z\rightarrow 0}f(z,\theta)=-\frac{2}{3}\left(3\cos^2\theta-1\right)(-1)^{1/4}\big[2K(-1)-iK(2)\big]\simeq -1.236\,\left(3\cos^2\theta-1\right),
\ee
with $K(z)$ being a complete elliptic function of first kind. Hence, one can analytically continue $f(z,\theta)$ to $z=0$, {\em defining} $f(0,\theta)=\lim_{z\rightarrow 0}f(z,\theta)$. Then, putting $z_0=0$ in (\ref{tauint1}) we have
\be\label{tauint2}
-\tau+c=\frac{a^{3/2}}{\sqrt{2M}}\big[f(z,\theta)-f(0,\theta)\big].
\ee
We now {\em promote} the integration constant $c$ to a new radial coordinate $\rho$: $c\rightarrow \rho$. Then
\be\label{rhodef}
\rho=\tau+\frac{a^{3/2}}{\sqrt{2M}}\big[f(z,\theta)-f(0,\theta)\big].
\ee
Differentiating (\ref{rhodef}) we get, for any {\em fixed} value of $\theta$
\be\label{rhodiff}
d\rho=d\tau+\frac{a^{3/2}}{\sqrt{2M}}\frac{df}{dz}\frac{dz}{dr}dr.
\ee
Since $\frac{dz}{dr}=\frac{1}{a}$ and
\be
\frac{df}{dz}=\frac{z^2+\cos^2\theta}{\sqrt{z(z^2+1)}},
\ee
[see (\ref{rdotint2})], we have
\be
d\rho=d\tau+\frac{1}{\sqrt{2M}}\frac{r^2+a^2\cos^2\theta}{\sqrt{r(r^2+a^2)}}dr,
\ee
or, equivalently
\be\label{eq1bis}
dr=\frac{\sqrt{A-\Delta\Sigma}}{\Sigma}(d\rho-d\tau).
\ee
Let us now introduce the following relationship
\be\label{eq2}
dt=\alpha d\rho + \beta d\tau,
\ee
with $\alpha$ and $\beta$ being free functions to be determined requiring (in mere analogy with the Schwarzschild case) that the resulting metric is (i) {\em synchronous} and (ii) {\em diagonal} in form.

Placing (\ref{eq1bis}) and (\ref{eq2}) in (\ref{bldiag}) we obtain
\be\label{dslem}
 ds^2=\frac{\Delta\Sigma}{A}(\alpha d\rho+\beta d\tau)^2-\Sigma d\theta^2-\frac{A-\Delta\Sigma}{\Delta\Sigma}(d\rho-d\tau)^2-\frac{A}{\Sigma}\sin^2\theta d\phi^2.
\ee
The above requirements imply
\bea\label{vincles}
\frac{\Delta\Sigma}{A}\beta^2-\frac{A-\Delta\Sigma}{\Delta\Sigma}=1,\\
\frac{\Delta\Sigma}{A}\alpha\beta +\frac{A-\Delta\Sigma}{\Delta\Sigma}=0.
\eea
We then find
\be\label{alphabetapm}
\beta=\pm\frac{A}{\Delta\Sigma},\quad\quad \alpha=\pm\bigg(1-\frac{A}{\Delta\Sigma}\bigg).
\ee
From (\ref{alphabetapm}) and (\ref{eq2}) we get [together with (\ref{eq1bis})] the following transformations
\be\label{transf1}
\left\{
	\begin{array}{ll}
dt= \pm\bigg(1-\frac{A}{\Delta\Sigma}\bigg)d\rho\pm\frac{A}{\Delta\Sigma}d\tau,\\
dr=\frac{\sqrt{A-\Delta\Sigma}}{\Sigma}(d\rho-d\tau).
\end{array}
\right.
\ee
In order to get the right Schwarzschild limit (\ref{transf}) when $a\rightarrow 0$, we need to choose the plus sign in the first equation (\ref{transf1}). Finally
\be\label{transfLem}
\left\{
	\begin{array}{ll}
dt= \bigg(1-\frac{A}{\Delta\Sigma}\bigg)d\rho+\frac{A}{\Delta\Sigma}d\tau,\\
dr=\frac{\sqrt{A-\Delta\Sigma}}{\Sigma}(d\rho-d\tau).
\end{array}
\right.
\ee
Substituting (\ref{transfLem}) in (\ref{bldiag}) again, we find
\be\label{ffzamo}
ds^2=d\tau^2-\bigg(1-\frac{\Delta\Sigma}{A}\bigg)d\rho^2-\Sigma d\theta^2-\frac{A}{\Sigma}\sin^2\theta d\phi'^2,
\ee
namely the Kerr metric in the (rotating) Lema\^itre coordinates. Since the employed rotating frame is the one pertaining to the family of Zero-Angular-Momentum-Observers (ZAMO) and the Lema\^itre coordinates are adapted to freely falling  observers, the coordinates in which (\ref{ffzamo}) has been written are adapted to an observer who is freely falling from the spatial infinity with zero initial velocity {\em and} zero angular momentum. These coordinates follows the observer's fall as well as the rotational dragging due to the Kerr geometry.

Notice that the parameters $A$, $\Delta$ and $\Sigma$ in (\ref{ffzamo}) depend on $\rho$ and $\tau$ through the radial function $r(\tau,\rho)$ which, in the general ($a\neq 0$) case, is implicitly given by (\ref{rhodef}).
 
%%%%%%%%%%%%%%%%%%%%%%%%%%%%%%%%    subsection 4.1     %%%%%%%%%%%%%%%%%%%%%%%%%%%%%%%%%%
\subsection{The Schwarzschild limit}
As a consistence check, let us show that the Schwarzschild form (\ref{lm}) is obtained as the limit $a\rightarrow 0$ of (\ref{ffzamo}). Putting $a=0$ in (\ref{ffzamo}) we find
\be\label{schwlimit}
ds^2=d\tau^2-\frac{2M}{r}d\rho^2-r^2d\theta^2-r^2\sin^2\theta d\phi'^2,
\ee
where it is implied that $r=r(\tau,\rho)$. This is indeed the ($\gamma=1$) Lema\^itre form (\ref{lm}). 
Let us also show that, in the limit $a\rightarrow 0$, the radial function $r(\tau,\rho)$ in (\ref{lm}) is given by (\ref{r}). 

Recall that $z=r/a$. Hence, $a\rightarrow 0$ implies $z\rightarrow \infty$. Looking at (\ref{fzeta}) we have
\be
\lim_{z\rightarrow \infty}f(z,\theta)=\lim_{z\rightarrow \infty} \frac{2}{3}\sqrt{z(z^2+1)},
\ee
so, in the limit $z\rightarrow\infty$, from (\ref{rhodef}) we get
\be
\rho=\tau+\frac{2}{3}\frac{a^{3/2}}{2M}z^{3/2},
\ee
and finally
\be
r=(2M)^{1/3}\bigg[\frac{3}{2}(\rho-\tau)\bigg]^{2/3}=r_g^{1/3}\bigg[\frac{3}{2}(\rho-\tau)\bigg]^{2/3},
\ee
namely, the explicit form (\ref{r}) of the radial function $r(\tau,\rho)$ in the Schwarzschild limit ($a=0$).

%%%%%%%%%%%%%%%%%%%%%%%%%%%%%%%%%    subsection 4.2     %%%%%%%%%%%%%%%%%%%%%%%%%%%%%%%%%%
\subsection{Constants of motion and Killing vectors}
Although the Lema\^itre form (\ref{ffzamo}) of the metric is explicitly dependent on the coordinate time $\tau$, we expect that the conserved quantities of the Kerr metric are still preserved. Indeed, a coordinate change cannot modify the dynamical properties of the considered spacetime. However, in the Lema\^itre coordinates, the Killing vector related to energy conservation is not $\partial_\tau$.
In oder to find the correct expression of the Killing vector related to energy consevation, let us consider the following transformation acting on the vectors $\partial_\tau$ and $\partial_\rho$ of the Lema\^itre coordinate basis
\bea
\left(
	\begin{array}{l}
	\partial_t \\
	\partial_r
	\end{array}
\right)=
{\cal{S}}^{-1}
\left(
	\begin{array}{l}
	\partial_\tau \\
	\partial_\rho
	\end{array}
\right),
\label{killingtransf}
\eea
where ${\cal{S}}$ is the matrix of the transformation (\ref{transfLem}). A straightforward calculation gives 
\be\label{killingLemaitre}
\partial_t=\partial_\tau+\partial_\rho.
\ee
Hence, recalling section 3.1, energy conservation in the Lema\^itre coordinates is related to the linear combination (\ref{killingLemaitre}), namely
\be\label{energyconsLemaitre}
\left(\delta^\mu_{(\tau)}+\delta^\mu_{(\rho)}\right)p_\mu=E.
\ee
In other words, $\xi^\mu=\delta^\mu_{(\tau)}+\delta^\mu_{(\rho)}$ is a Killing vector for the Kerr spacetime.

%%%%%%%%%%%%%%%%%%%%%%%%%%%%%%%%%%%%    section 5     %%%%%%%%%%%%%%%%%%%%%%%%%%%%%%%%%%
\section{Horizons and observers in the Lema\^itre coordinates}

The Kerr metric (\ref{ffzamo}) can be rewritten as
\be\label{ffzamo2}
ds^2=d\tau^2-\frac{2Mr(r^2+a^2)}{A}d\rho^2-\Sigma d\theta^2-\frac{A}{\Sigma}\sin^2\theta d\phi^2.
\ee
We immediately see the absence of coordinate singularities at the horizons as those found in the Boyer-Lindquist coordinates (the roots of $\Delta=0$). Only a non-removable, physical (ring) singularity remains, at $r=0$, $\theta=\pi/2$. Nevertheless, the surfaces at $r=r_\pm$, identifying the event and Cauchy horizons of the Kerr black hole still play a physical role, as we will see below.

%%%%%%%%%%%%%%%%%%%%%%%%%%%%%%%%    subsection 5.1     %%%%%%%%%%%%%%%%%%%%%%%%%%%%%%%%%%
\subsection{Horizons}

Due to their very construction, Lema\^itre coordinates follow the freely falling observers in the Kerr metric. As a consequence, in those coordinates the geometrical surfaces defining the Kerr horizons become {\em dynamical} geometrical objects evolving in the observer's proper time. Thanks to the axisymmetry of the Kerr spacetime and its reflection symmetry with respect to the equatorial plane $\theta=\pi/2$, we may discard the azimuthal coordinate $\phi$, hence describing  the horizon solely by means of the $(r,\theta)$ Boyer-Lindquist coordinates, with $0\leq r<+\infty$ and $0\leq\theta\leq \pi/2$. Using (\ref{rhodef}) with $z=z_{\pm}$, we get
\be\label{horevolution}
\tau=\rho-\frac{a^{3/2}}{\sqrt{2M}}\big[f(z_{\pm},\theta)-f(0,\theta)\big],
\ee
where $z_{\pm}=r_{\pm}/a$, ($a\neq 0$) are the adimensional radial coordinates of the horizons (\ref{horizon}) in the Boyer-Lindquist  coordinates.
(\ref{horevolution}) gives the evolution of the Kerr horizons ($r_{\pm}$) in the Lema\^itre coordinates.

%%%%%%%%%%%%%%%%%%%%%%%%%%%%%%%%    subsection 5.2     %%%%%%%%%%%%%%%%%%%%%%%%%%%%%%%%%%
\subsection{Lema\^itre observers}

In the Lema\^itre coordinates, freely falling observers are {\em at rest}. Each observer {\bf u} is identified by means of fixed values of the coordinates $\rho$ and $\theta$: {\bf u}$(\ovl\rho,\ovl\theta)$ \footnote{More properly, each pair $(\ovl\rho,\ovl\theta)$ represents the {\em family} of all the observers differing for the ininfluent azimuthal coordinate $\phi$.}.

All the observers {\bf u}$(0,\theta)$ meet the point $r=0$ at the time $\tau=0$. Observers having $\ovl\rho>0$ will meet $r=0$ at $\tau>0$. So, these latter observers {\em follow in time} - so to say - the observers {\bf u}$(0,\theta)$.

In a "reduced" coordinate frame $(\rho,\,\tau)$, the worldline of a given Lema\^itre observer {\bf u}$(\ovl\rho,\ovl\theta)$, whose equations are $\rho=\ovl\rho$, $\theta=\ovl\theta$, appears as a vertical straight line,  parallel to the time axis. Observers sharing the same value $\ovl\rho$ and different $\theta$, will meet the "point" $r=0$ at the same time:
\be
\tau=\ovl\rho-\frac{a^{3/2}}{\sqrt{2M}}\big[f(0,\theta)-f(0,\theta)\big]=\ovl\rho,\quad\quad\forall\,\, \theta.
\ee
On the other hand, those observers will meet the horizons at {\em different} times, depending on the value of $\ovl\theta$. Actually, putting $\rho=\ovl\rho$ in (\ref{horevolution}) we find
\be
\tau_{\pm}=\ovl\rho-\frac{a^{3/2}}{\sqrt{2M}}\big[f(z_{\pm},\theta)-f(0,\theta)\big],
\ee
The lapse of time required to cross the inter-horizon region depends on the observer colatitude $\ovl\theta$ only, no matter what the value of $\rho$ is. We have
\be\label{deltataulapse}
\Delta\tau=\tau_--\tau_+=\frac{a^{3/2}}{\sqrt{2M}}\big[f(z_+,\ovl\theta)-f(z_-,\ovl\theta)\big],\quad\quad\forall\,\, \rho.
\ee
\begin{figure}[ht]
\begin{center}
\includegraphics[angle=0,width=.60\textwidth]{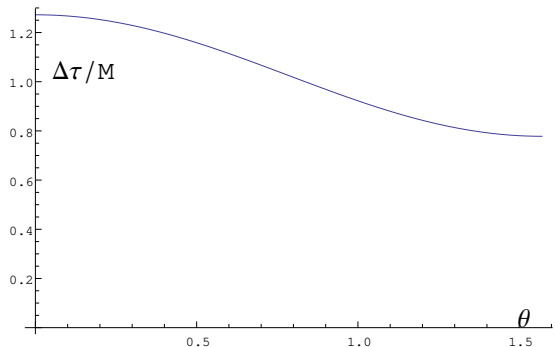}%
\caption{Plot of the ratio $\Delta\tau/M$ as a function of the colatitude $\theta$ [see (\ref{deltataulapse})] for a Kerr black hole with $a/M=1/\sqrt{2}$. The values of the adimensional coordinate $z$ corresponding to the horizons $r_{\pm}$ are $z_{\pm}=\sqrt{2}\pm 1$.   Equatorial observers experience a shorter inter-horizon crossing time than the polar observers.  \label{fig1}}
\end{center}
\end{figure}
This is an obvious consequence of the time-translation invariance of the Kerr spacetime.
We see that $\Delta\tau$ decreases as $\theta$ increases. Hence equatorial observers experience a shorter inter-horizon crossing time than the polar observers. In Fig.1 a plot of the adimensional ratio $\Delta\tau/M$ as a function of the observer's latitude $\theta$ is shown.

%%%%%%%%%%%%%%%%%%%%%%%%%%%%%%%%   subsection 5.3     %%%%%%%%%%%%%%%%%%%%%%%%%%%%%%%%%%
\subsection{Introducing an adapted tetrad}

Looking at the metric (\ref{ffzamo}), the construction of a tetrad frame $\lambda^\mu_{\hat a}$, adapted to the  Lema\^itre observers is straightforward. We find
\bea
\left\{
	\begin{array}{ll}
	\lambda_{\hat t}=\partial_\tau,\\
	\lambda_{\hat x}=\bigg(1-\frac{\Delta\Sigma}{A}\bigg)^{-1/2}\partial_\rho,\\
	\lambda_{\hat y}=\frac{1}{\sqrt{\Sigma}}\partial_\theta,\\
	\lambda_{\hat z}=\frac{\sqrt{\Sigma}}{\sqrt{A}\sin\theta}\partial_\phi,\\
	\end{array}
\right.
\label{tetrad}
\eea
where it is understood that the parameters $A$, $\Delta$ and $\Sigma$ depend on $\rho$ and $\tau$ through the radial function $r(\rho,\tau)$ implicitly given by (\ref{rhodef}). Thus, each fixed pair of values $(\rho,\theta)$, specifies the tetrad (\ref{tetrad}) adapted to the Lema\^itre observer ${\bf u}(\rho,\theta)$.

It is an easy task to prove that the above tetrad is geodesic (obviously) and Fermi-Walker transported \cite{bini,fdf}. Actually, the Fermi coefficients $\Lambda_{\hat a\hat b}=\lambda_{\hat a\mu;\nu}\lambda_{\hat b}^\mu\lambda_{\hat t}^\nu$ are identically vanishing:
\be
\Lambda_{\hat x\hat y}=\Lambda_{\hat x\hat z}=\Lambda_{\hat y\hat z}=0.
\ee
In that respect, the tetrad frame (\ref{tetrad}) can be regarded as the best approximation to an inertial reference frame in the Kerr geometry.

%%%%%%%%%%%%%%%%%%%%%%%%%%%%%%%%    section 6     %%%%%%%%%%%%%%%%%%%%%%%%%%%%%%%%%%
\section{A journey towards the Cauchy horizon}

In this section we will apply the above results discussing the measurements performed by a couple of observers - say, Alice and Bob - freely falling into  a Kerr black hole.

%%%%%%%%%%%%%%%%%%%%%%%%%%%%%%%%    subsection 6.1     %%%%%%%%%%%%%%%%%%%%%%%%%%%%%%%%%%
\subsection{Geodesic of radial photons}
Without loss of generality, we may assume that Alice has $\rho=\rho_A=0$. Bob, having $\rho=\rho_B>0$, follows Alice in her fall into the black hole along the same path (this means that Alice and Bob share the same angular coordinates $\theta$ and $\phi$). Suppose that Alice communicates with Bob, sending radial photons, outwards with respect to the black hole. Such photons have $d\theta=0$, $d\phi=0$ and $ds=0$. From (\ref{ffzamo}) we have
\be\label{photon1}
\frac{d\tau}{d\rho}=\pm\left(\frac{A-\Sigma\Delta}{A}\right)^{\frac{1}{2}}\equiv\pm\Phi(z,\theta).
\ee
Being interested in the outgoing photons, we will consider the $+$ sign (the $-$ sign refers to the ingoing photons).
We immediately see that $\Phi>1$ in the inter-horizon region $\Delta<0$, while $\Phi=1$ on the horizons $z_{\pm}$. Hence, outgoing photons, emitted by Alice at $\tau_\pm$ (i.e., when Alice meets the outer and the inner horizon) will stay forever on the $z_{\pm}$ horizons. 
Let us rewrite (\ref{eq1bis}) as
\be\label{drphoton}
dr=\frac{\sqrt{A-\Delta\Sigma}}{\Sigma}\left(\frac{d\rho}{d\tau}-1\right)d\tau=\pm\frac{\sqrt{A}}{\Sigma}\left(1\mp\sqrt{1-\frac{\Delta\Sigma}{A}}\right)d\tau,
\ee
\begin{figure}[ht]
\begin{center}
\includegraphics[angle=0,width=.40\textwidth]{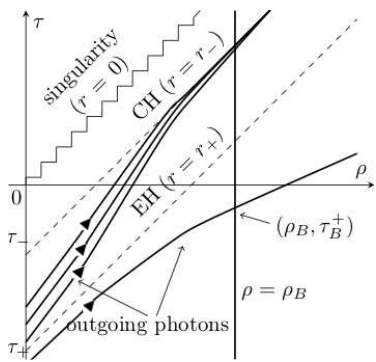}%
\caption{Schematic drawing of photon worldlines emitted by Alice  (whose worldline is the time axis, $\rho=\rho_A=0$) during her inter-horizon journey. Alice crosses the event horizon (EH) at time $\tau_+$ and reaches the Cauchy horizon (CH) at $\tau_-$. Photons are emitted in the outward radial direction, on the Kerr equatorial plane. A photon emitted before $\tau_+$ flies away towards the spatial infinity, and can be detected by Bob at $(\rho_b,\tau^+_B)$. Photons emitted at a time $\tau\in(\tau_+,\tau-)$ will pile up on the Cauchy horizon. A late observer (Bob, $\rho=\rho_B$) will receive such photons in a very short lapse of time, in which almost all the story of Alice's inter-horizon journey is condensed.  Arbitrary units have been employed. \label{fig2}}
\end{center}
\end{figure}
where use of (\ref{photon1}) has been made. In (\ref{drphoton}) the upper (lower) sign refers to radially outgoing (ingoing) photons. Looking at the outgoing photons, we see that, as time goes on ($d\tau>0$),
\bea\label{fly}
dr>0 \quad {\rm if}\quad \Delta>0,\nonumber\\
dr<0 \quad {\rm if}\quad \Delta<0.
\eea
Hence, photons emitted by Alice at a time $\tau<\tau_+$ (i.e., before crossing the event horizon $r=r_+$), will fly away, towards the spatial infinity ($r\rightarrow \infty$). Bob can indeed detect such photons at some late time, before he reaches the event horizon at time $\tau^+_B=\rho_B+\tau_+$ (see Fig. 2).
On the other hand, outgoing photons emitted by Alice at a time $\tau_+<\tau<\tau_-$, having worldline slopes greater than unity,  will asymptotically end on the inner horizon. The Cauchy horizon will eventually store all those photons. Bob will receive such photons during his inter-horizon journey. As photons pile up on the Cauchy horizon, Bob  will see the Alice clock to speed up meanwhile he approaches the Cauchy horizon. Hence, according to Bob, almost all of Alice's journey between the horizons will be condensed in a very short lapse of time (see Fig.2).

%%%%%%%%%%%%%%%%%%%%%%%%%%%%%%%%%    subsection 6.2     %%%%%%%%%%%%%%%%%%%%%%%%%%%%%%%%%%
\subsection{A numerical example}
The photon worldlines roughly plotted in Fig.2 can be obtained by means of the corresponding parametric equations. 
Solving (\ref{drphoton}) with respect to $d\tau$ and recalling that $dr=adz$ we find
\be\label{detauzeta}
d\tau=\pm\frac{a\Sigma}{\sqrt{A}}\left(1\mp\sqrt{\frac{A-\Sigma\Delta}{A}}\right)^{-1}dz.
\ee
Solving (\ref{photon1}) with respect to  $d\rho$ and using (\ref{detauzeta}), we also get
\be\label{derhozeta} 
d\rho=\frac{a\Sigma}{\sqrt{A-\Sigma\Delta}}\left(1\mp\sqrt{\frac{A-\Sigma\Delta}{A}}\right)^{-1}dz.
\ee
Expressing (\ref{detauzeta}) and (\ref{derhozeta}) as functions of $z$, we obtain
\bea\label{detauzetaz}
 d\tau=\pm\frac{M\alpha^{3/2}\left(z^2+\cos^2\theta\right)}{\left(\alpha (z^2+1)(z^2+\cos^2\theta)+2z\sin^2\theta\right)^{1/2}}\nonumber\\
\times \left(1\mp\sqrt{\frac{2z(z^2+1)}{\alpha (z^2+1)(z^2+\cos^2\theta)+2z\sin^2\theta}}\right)^{-1}dz.
\eea
and
\be\label{derhozetaz} 
 d\rho=M\alpha^{3/2}\frac{\left(z^2+\cos^2\theta\right)}{\sqrt{2z(z^2+1)}}\left(1\mp\sqrt{\frac{2z(z^2+1)}{\alpha (z^2+1)(z^2+\cos^2\theta)+2z\sin^2\theta}}\right)^{-1}dz.
\ee
Upon integration, (\ref{detauzetaz}) and (\ref{derhozetaz}) yield the {\em parametric} equations (with respect to the parameter $z$) of the worldlines of the radial photons (the upper sign refers to outgoing photons, the lower sign to ingoing photons).

Consider, as an example, the observer {\bf u}$(0,\pi/2)$ in the equatorial plane. Assuming, as before, $\alpha=a/M=\frac{1}{\sqrt{2}}$, we have, for the outgoing photons
 \be
 d\tau=2^{-\frac{1}{2}} M\frac{\sqrt{\frac{z^3}{z^3+z+2\sqrt{2}}}}{1-2^{3/4} \sqrt{\frac{z^2+1}{z^3+z+2\sqrt{2}}}}dz.
 \ee
 \be
d\rho=2^{-\frac{5}{4}}M\frac{\frac{z^2}{\sqrt{z\left(1+z^2\right)}}}{1-2^{3/4} \sqrt{\frac{\left(z^2+1\right)}{z^3+z+2\sqrt{2}}}}dz,
\ee
Numerically integrating and plotting the solution we obtain the result shown in Fig.3.
\begin{figure}[ht]
\begin{center}
\includegraphics[angle=0,width=.70\textwidth]{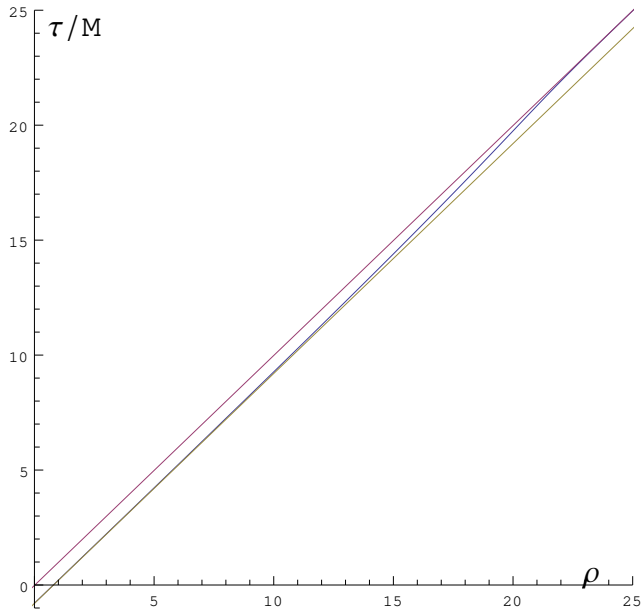}%
\caption{Plot of the numerical output for an outgoing photon wordline (the curved line) in the Lema\^itre coordinates $(\rho,\,\tau)$, in the case of a  Kerr black hole having $a/M=1/\sqrt{2}$. The upper and lower straight lines represent  the Cauchy horizon (CH) and the event horizon (EH), respectively. The photon is emitted by the $\rho=0$ observer just after the EH crossing ($\tau\simeq -.79\,M$) and reaches the CH after a lapse of time $\sim\Delta\tau\simeq 26\,M$. \label{fig3}}
\end{center}
\end{figure}
The two straight lines with a $\pi/4$ slope represent the worldlines of the Cauchy (upper line) and event (lower line) horizon in the $(\rho,\tau)$ Lema\^itre coordinates. The curve between the two straight lines represents the worldline of an outgoing photon, emitted by the  {\bf u}$(0,\pi/2)$ observer at $\tau\simeq -.79\,M$  just after crossing the event horizon ($\tau_+\simeq  .79\,M$). The photon worldline approaches the Cauchy horizon after a lapse of time $\Delta\tau\simeq 26\,M$, while the observer meets the Cauchy horizon in a shorter lapse of time, $\Delta\tau\sim .78\,M$.

%%%%%%%%%%%%%%%%%%%%%%%%%%%%%%%%%%5    section 7     %%%%%%%%%%%%%%%%%%%%%%%%%%%%%%%%%%
\section{Photon blue-shift near the Cauchy horizon}
As we pointed out above, Bob sees Alice's clock to speed up as he approaches the Cauchy horizon. This is closely related to the photon blue-shift near $r=r_-$. 

Consider a photon, emitted by an observer A (Alice) $\rho=\rho_A$ at a time $\tau=\tau_e>\tau_+$, i.e., after the event horizon crossing. Since the spacetime admits a Killing vector $\xi^\mu=\delta^\mu_{(\tau)}+\delta^\mu_{(\rho)}$ [see (\ref{energyconsLemaitre})], we have
\be\label{Econst}
\xi^\mu k_\mu={\cal E}={\rm const},
\ee
along the null photon geodesic, whose tangent vector $k^\mu$ is the photon wavevector, obeying $k^\mu k_\mu=0$. The geodesic is parametrized by an affine parameter $\sigma$, so that $k^\tau=\frac{d\tau}{d\sigma}=\omega_e$, where $\omega_e$ is the photon frequency as {\em measured} by Alice $u^\mu=\delta^\mu_\tau$ at the emission point, namely: $\omega_e=u^\mu k_\mu=k_\tau=g_{\tau\tau}k^\tau=k^\tau$. Furthermore, $k^\rho=\frac{d\rho}{d\sigma}=\frac{d\rho}{d\tau}\frac{d\tau}{d\sigma}=\frac{d\rho}{d\tau}\omega_e$. From (\ref{Econst}) we get
\be\label{Econst1}
\left(\omega+g_{\rho\rho}\omega\frac{d\rho}{d\tau}\right)_e={\cal E},
\ee
where the subscript $e$ means that all the quantities in round brackets have to be evaluated at the emission point.
Assuming a radial $(d\theta=d\phi=0)$ {\em outgoing} photon, from (\ref{ffzamo}) we have 
\be\label{outphoton}
\frac{d\rho}{d\tau}=\frac{1}{\sqrt{-g_{\rho\rho}}}=\frac{1}{\sqrt{1-\frac{\Delta\Sigma}{A}}}. 
\ee
Combining (\ref{Econst1}) and (\ref{outphoton}) we find
\be\label{Econst2}
\omega_e\left(1-\sqrt{1-\frac{\Delta\Sigma}{A}}\right)_e={\cal E}={\rm const}
\ee
along the photon geodesic. Recall that outgoing photons emitted in the inter-horizon region fly towards the Cauchy horizon [see the discussion below (\ref{fly})].
Consider now a second observer B (Bob), who receives the photon at $r=r_r<r_e$. Let $\omega_r$ be the photon frequency measured by Bob. Then
\be\label{Econst3}
\omega_r\left(1-\sqrt{1-\frac{\Delta\Sigma}{A}}\right)_r=\omega_e\left(1-\sqrt{1-\frac{\Delta\Sigma}{A}}\right)_e.
\ee
Hence
\be\label{omegashift}
\omega_r=\frac{\left(1-\sqrt{1-\frac{\Delta\Sigma}{A}}\right)_e}{\left(1-\sqrt{1-\frac{\Delta\Sigma}{A}}\right)_r}\,\omega_e.
\ee
Notice that in the inter-horizon zone ($\Delta<0$) both the quantities in round brackets are $<0$. If Bob is nearby the Cauchy horizon, then $r_r\rightarrow r_-$, and $\Delta\rightarrow 0^-$ in the denominator of (\ref{omegashift}). Hence
\be\label{blueinfty}
\omega_r\rightarrow\infty,
\ee
for photons detected nearby the Cauchy horizon. This is a rather well-known effect, closely related to the (open) issues concerning the Cauchy horizon instability \cite{chandra2,toporenski}.

%%%%%%%%%%%%%%%%%%%%%%%%%%%%%%%%%%%    section 8     %%%%%%%%%%%%%%%%%%%%%%%%%%%%%%%%%%
\section{Concluding remarks}

In this paper we have obtained a transformation from the Boyer-Lindquist coordinates to a new coordinate set, adapted to a family of freely falling observers in the Kerr geometry.

We named such new coordinates {\em Lema\^itre coordinates}, since they generalize the well-known ones originally introduced by Lema\^itre in the Schwarzschild geometry. In these Lema\^itre coordinates the Kerr metric takes a diagonal and syncronous form (i.e., the lapse funcion is unity). 
Furthermore the metric becomes free from coordinate singularities at the Kerr horizons. The natural tetrad field adapted to the Lema\^itre (i.e., freely falling) observers is geodesic and Fermi-Walker transported, hence providing the best approximation to an inertial frame in a curved spacetime.

Being free from coordinate singularities, the proposed Lema\^itre coordinates are horizon penetrating, hence offering a potential tool to explore the spacetime region behind the event horizon, meanwhile adopting the physical point of view of an observer in free fall.
 
As an application, we have studied the behaviour of radial photons, showing the existence of a blue-shift effect as measured by a Lema\^itre observer close to the Cauchy horizon. 
Such a result fairly agrees with other similar findings appeared in the literature \cite{chandra2,toporenski} and can be related to some open issues concerning the Cauchy horizon (in)stability, as briefly discussed in the intoduction.

Being synchronous, the Lema\^itre form of the Kerr metric is also suitable for a Hamiltonian formulation of a quantum field theory \cite{fulling}.
In that respect, the proposed form of the Kerr metric might prove useful to investigate the behaviour of quantum fields near and behind the event horizon, down to the Cauchy horizon. Such analysis can be performed both analytically, following a near-horizon approach (see, e.g., \cite{camblong,azizi}), as well as  by means of numerical computations (see section 6.2). This will be the goal of further investigation.

%\section*{Acknowledgments}

\end{document}